\documentclass[aps,prb,twocolumn,superscriptaddress]{revtex4-1}

\usepackage{color}
\usepackage{times}
\usepackage{epsfig}
\usepackage{amsmath}
\usepackage{amssymb}

\def\bea{\begin{eqnarray}}      \def\eea{\end{eqnarray}}

\def\bv {\boldsymbol{\nu}}

\begin{document}

\title{Majorana zero modes in dislocations of ${\rm Sr_2RuO_4}$}

\author{Taylor L. Hughes}
\affiliation{Department of Physics and Institute for Condensed Matter Theory, University of Illinois at Urbana-Champaign, Urbana IL 61801, USA}
\author{Hong Yao}
\affiliation{Institute for Advanced Study, Tsinghua University, Beijing, 100084, China}
\author{Xiao-Liang Qi}
\affiliation{Department of Physics, Stanford University, Stanford, CA 94305, USA}

\date{\today}

\begin{abstract}
We study the topologically protected Majorana zero modes induced by lattice dislocations in chiral topological superconductors. Dislocations provide a new way to realize Majorana modes at zero magnetic field. In particular, we study several different types of dislocations in the candidate material ${\rm Sr_2RuO_4}.$ We also discuss the properties of linked dislocation lines and linked dislocation and flux lines. Various experimental consequences are predicted which provide a new approach to determine the nature of the superconducting phase of ${\rm Sr_2RuO_4}.$
\end{abstract}

\maketitle

Chiral topological superconductors (TSC) in 2D are predicted to have vortices that harbor stable Majorana bound states (MBS) when the Chern number topological invariant is an odd integer\cite{read2000}. This property is an important feature for  topological quantum computing architectures which are based on the use of such non-Abelian anyon qubits\cite{kitaev2003,nayak2008}. It is generally agreed that Sr$_2$RuO$_4$ is a quasi-2D, p-wave superconductor with broken time-reversal symmetry, although the precise nature of the order parameter is still controversial (for a review, see Ref. \onlinecite{kallin2012}). 
So, while there is no comprehensive evidence that Sr$_2$RuO$_4$ is a TSC, it is one of the best candidate chiral TSC materials.

The low-energy electronic structure of the normal metal state of Sr$_2$RuO$_4$ is controlled by the $t_{2g}$ multiplet of d-orbitals $d_{xz}$, $d_{yz}$, and $d_{xy}$. These three orbitals give rise to three Fermi-surfaces which are expected to become fully gapped below the superconducting transition temperature at $T_c\sim 1.5K.$ The conventional wisdom indicates that the quasi-2D $d_{xy}$ band dominates the pairing instability and develops a nodeless chiral $p_x +i p_y$ order parameter\cite{rice1995}. If such an order parameter were generated, then the recently measured half-quantum vortices would indeed bind MBS\cite{leggett-rmp,raffi2011}. However, a conflict between the theoretical prediction of chiral surface states in the $p_x+ip_y$ state, and the clear lack of surface currents measured in Ref. \onlinecite{kam2005}, motivated Raghu, Kapitulnik, and Kivelson (RKK) to make a different,  compelling proposal for the nature of the order parameter which is dominated instead by the quasi-1D $d_{xz}$ and $d_{yz}$ orbitals\cite{rkk2010}. The RKK order parameter by itself is odd-parity/p-wave which breaks time-reversal, but is not a chiral TSC and should not exhibit chiral edge states. Thus, it is consistent with the measurements of Ref. \onlinecite{kam2005}. The true nature of the order parameter is still an open question, and experiments that can distinguish between these two predictions are needed.

In this article we propose topological properties that can distinguish the two pairing schemes and also provide a new approach to MBS in superconductors with even-integer or vanishing Chern number. Our key observation is that the RKK order parameter, while trivial in the sense of a 2D chiral superconductor (i.e. vanishing Chern number), in fact is still nontrivial as a \emph{weak} TSC. Weak TSCs (and insulators)\cite{fu2007b,moore2007,roy2009a,kitaev2009} 
are topological states protected by translation symmetry. They are distinguished by a topological invariant defined in a 
lower-dimensional sub-manifold of the Brillouin zone (BZ) (recall that the so-called \emph{strong} invariant depends on the electronic structure in the entire BZ). For example, the 3D weak topological insulators are classified by three $Z_2$ invariants, which characterize whether the gapped Bloch Hamiltonian restricted to the three 2D planes of $k_x=\pi,k_y=\pi,k_z=\pi$ in the BZ are trivial insulators or quantum spin Hall insulators in that plane. Heuristically, a non-trivial weak invariant indicates a state which is made from stacking up topological states of lower dimensions. Another example is a 2D superconductor in class D (no symmetry) which has two weak $Z_2$ indices. These exist since there is a strong $Z_2$ topological superconductor in this class in 1D --the Majorana chain/p-wave wire\cite{kitaev2001,schnyder2008,kitaev2009}. We show that the RKK pairing in the $d_{xz}$ and $d_{yz}$ bands corresponds to this type of weak topological superconductor where 1D topological wires have been ``stacked" into a higher dimension. As a consequence,
we show that naturally occurring or fabricated crystal defects can exhibit a number of remarkable properties that can help to distinguish the case when the $p_x+ip+y$ state dominates from the case when the RKK pairing dominates. These properties can topologically characterize the superconducting order, and as we will show, can give rise to a mechanism for stable MBS even in the RKK state.

\begin{figure}[t]
    \begin{center}
        \includegraphics[width=3.2in]{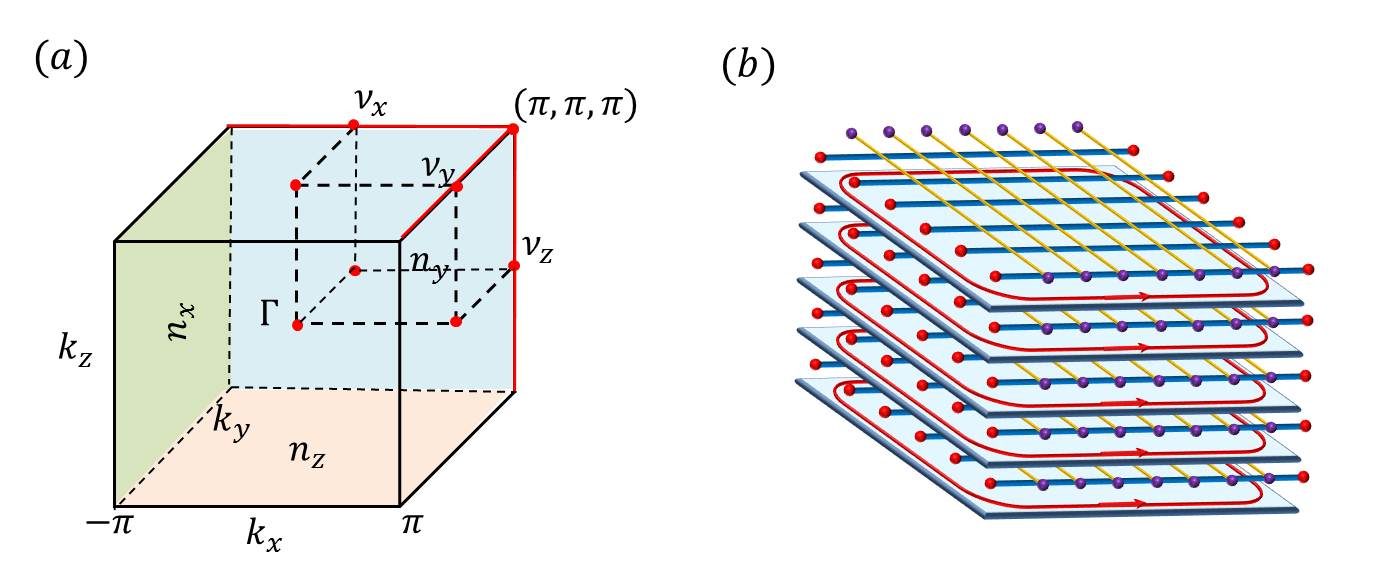}
    \end{center}
    \caption{(a) The Brillouin zone of a 3D SC. The primary weak topological invariants ${\bf n}=(n_x,n_y,n_z)$ are defined as the Chern numbers in the three independent planes (colored green, blue and red), and the secondary weak topological invariants $\bv=(\nu_x,\nu_y,\nu_z)$ are defined as the 1D ${\rm Z_2}$ indices along the three perpendicular lines colored in red. (b) Illustration that a topological superconductor with topological invariants ${\bf n}=(0,0,1),~\bv=(1,1,0)$, such as ${\rm Sr_2RuO_4}$ is equivalent to decoupled layers of chiral superconductors (with the red line and arrow labeling the chiral edge states) and 1D wires along $x$ and $y$ directions with Majorana zero modes at end points. }
    \label{fig1}
\end{figure}

\emph{3D \emph{weak} topological superconductors}:
The non-interacting topological insulators and superconductors in generic dimensions have been classified\cite{kitaev2009,schnyder2008,qi2008b}. Here we are interested in superconductors in class D, which have a strong $Z_2$ topological classification in 1D, $Z$ in 2D, and trivial in 3D which is stable without any additional symmetries beyond fermion parity conservation. For class D SCs in 3D with additional translational symmetries we can define weak topological invariants as well---the 2D Chern numbers  can be defined along constant $k_x,k_y$ or $k_z$ planes in the BZ. In a gapped state the Chern number cannot change, so the Chern number in different $k_z=\rm{const}$ planes is the same integer $n_z.$ Similarly $n_x$ and $n_y$ can be defined for the other two planes, as is illustrated in Fig. \ref{fig1}(a). The integer-valued vector ${\bf n}=(n_x,n_y,n_z)$ are the {\it primary weak indices} of the 3D TSC. A system in class D with indices ${\bf n}$ is \emph{topologically} equivalent to a set of decoupled 2D layers of topological chiral superconductors with non-vanishing Chern number, stacked along the ${\bf n}$ direction. For any surface plane which is not perpendicular to ${\bf n}$, there will be chiral surface states.

Similarly,  the 1D ${\rm Z_2}$ invariants\cite{kitaev2001} 
can be calculated along time-reversal invariant lines in the 3D BZ\footnote{The 1D ${\rm Z_2}$ topological invariants are defined only when the particle-hole symmetry is respected, which is guaranteed for the Bloch (Bogoliubov de-Gennes) Hamiltonians defined along time-reversal invariant lines in the 3D BZ.}. The three \emph{secondary weak topological invariants} are defined as the $Z_2$ invariants along the three lines  $(k_y,k_z)=(\pi,\pi),~(k_z,k_x)=(\pi,\pi),~(k_x,k_y)=(\pi,\pi).$ We collect them into a vector $\bv=(\nu_x,\nu_y,\nu_z)$ shown in Fig. \ref{fig1} (a) (with $\nu_{x,y,z}=0,1$). It can be shown that $\bv$ and ${\bf n}$ together determine the $Z_2$ invariants along all other time-reversal invariant lines. A TSC with $\bv\neq 0$ and ${\bf n}=0$ is \emph{topologically} equivalent to decoupled 1D TSC wires aligned in the direction of $\bv$, each of which has an odd number of MBS at each end. Consequently, for any surface plane that is not \emph{parallel} to ${\bf \nu}$ there will be Majorana surface states. A generic TSC with both ${\bf n}$ and $\bv$ non-vanishing can be considered as decoupled layers of 2D chiral TSC coexisting with decoupled wires of 1D TSC.  We will show that the RKK model (ignoring spin degeneracy) has the  weak topological invariants ${\bf n}=(0,0,1),~\bv=(1,1,0)$, which is topologically equivalent to decoupled topological layers and wires as is illustrated in Fig. \ref{fig1} (b). This conceptual decomposition into stacks of lower dimensional systems will be helpful to illustrate our discussion of dislocations. Primary weak topological indices were first discussed in the context of time-reversal invariant topological insulators\cite{fu2007b,moore2007,roy2009} and subsequently both primary and secondary indices (and beyond) can be straightforwardly extracted for the entire periodic table of topological states from the K-theory calculation in Ref. \onlinecite{kitaev2009}. We note that although we have defined the secondary-index $\bv$ as a vector, the natural structure is actually an anti-symmetric two-index tensor which can be interpreted as a vector only in 3D.\cite{ran2010,teo2010} 
In addition to the application to Sr$_2$RuO$_4$, one of our primary results is that secondary weak invariants also describe the MBS trapped on a pair of \emph{linked} dislocation lines, which is analogous to a similar mechanism for bound states on linked vortex lines in 3D time-reversal invariant topological superconductors\cite{qi2009b}.

\begin{figure}[t]
    \begin{center}
        \includegraphics[width=1.4in]{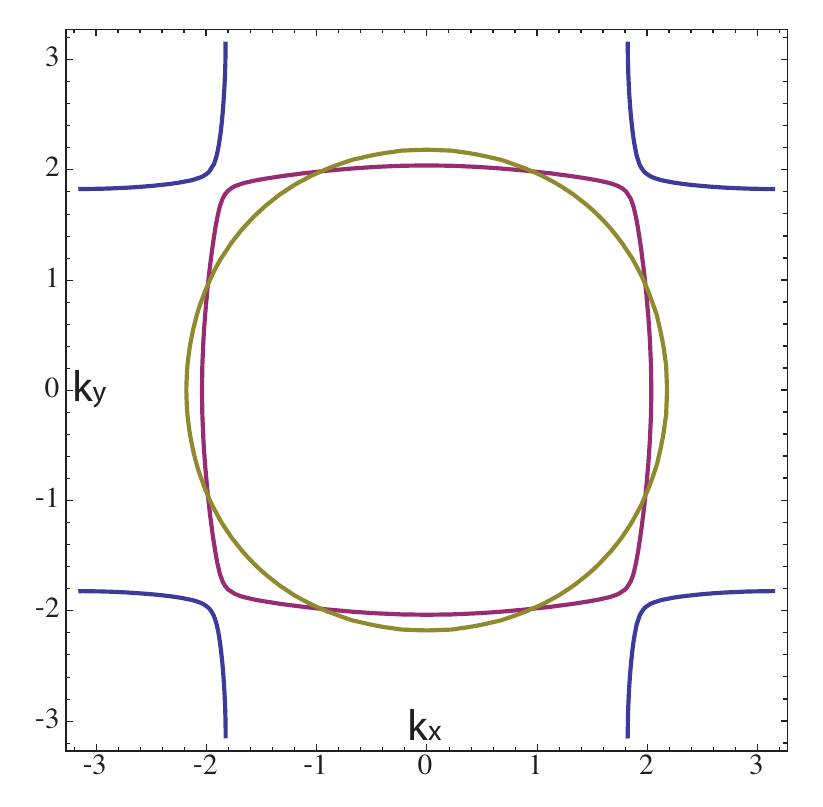}\includegraphics[width=1.8in]{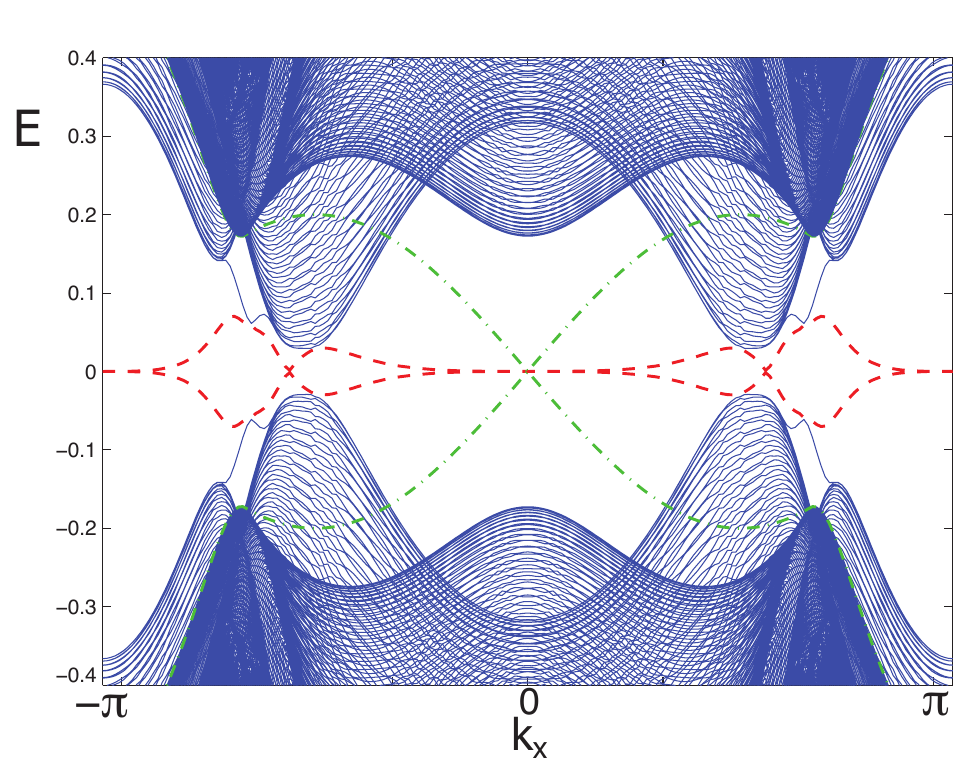}
    \end{center}
    \caption{(Left) Fermi-surface structure of Sr$_2$RuO$_4$ showing the Fermi surfaces coming from the d$_{xz}$ d$_{yz}$ and d$_{xy}$ bands with orbital mixing.  (Right) Low-energy quasi-particle spectrum of superconducting Sr$_2$RuO$_4$ for a geometry with open boundary conditions in the $y$-direction and periodic boundary conditions on the $x$-direction. The states in red-dashed lines are edge states arising from the quasi-1D bands; notice they exist at $k_x=0$ and $k_x=\pi.$ The green dash-dotted lines are the edge states arising from the quasi-2D band; notice they exist only around $k_x=0.$}
    \label{fig:edgestates}
\end{figure}

\emph{Application in ${\rm Sr_2RuO_4}$}:
To begin our discussion we will review the electronic structure of the normal metal state of Sr$_2$RuO$_4.$ This will by followed by recounting the superconducting pairing 
scheme as given in the paper by 
RKK\cite{rkk2010}. 
The three relevant orbitals for the electronic structure are the $t_{2g}$ multiplet $d_{xz},d_{yz}, d_{xy}$ which will be labelled by $\alpha=1,2,3.$ The layered structure of Sr$_2$RuO$_4$ makes the system behave quasi-two dimensionally; consequently the first two orbitals are effectively quasi-1D in nature while $d_{xy}$ is quasi-2D. The bandstructure can be modeled using these three orbitals, plus spin, on a 
simple tetragonal lattice with nearest neighbor, and next-nearest neighbor hoppings. The Bloch Hamiltonian  is
\begin{eqnarray}
H(\bf k)&=&\left(\begin{array}{ccc}\epsilon_{xz}(\bf k) & \Lambda(\bf k) & 0\\ \Lambda(\bf k)& \epsilon_{yz}(\bf k)  & 0 \\ 0 & 0 &\epsilon_{xy}(\bf k) \end{array}\right)\otimes \mathbb{I}_{spin}\\
\epsilon_{xz}(\bf k)&=&-2t\cos k_x -2t^{\perp}\cos k_y - 2t^z_1 \cos k_z\nonumber\\
\epsilon_{yz}(\bf k)&=&-2t\cos k_y -2t^{\perp}\cos k_x-2t^z_1 \cos k_z\nonumber\\
\epsilon_{xy}(\bf k)&=&-2t'(\cos k_x+\cos k_y)-4t'' \cos k_x \cos k_y-2t^z_2 \cos k_z\nonumber\\
\Lambda(\bf k)&=&-2\lambda\sin k_x \sin k_y\nonumber
\end{eqnarray}\noindent where values of in-plane hopping parameters taken from RKK are $t=1.0, t'=0.8, t^{\perp}=0.1, t''=0.3$. We have also 
considered 
an orbital-hybridization term $\Lambda(\bf k)$ which arises from next-nearest neighbor hopping between different quasi-1D orbitals in the $xy$-plane, and removes the crossings in those Fermi surfaces. 
Hopping amplitudes along $z$ are $t^z_1$ and $t^z_2$ for quasi-1D and quasi-2D orbitals, respectively. Due to the layered structure of the lattice, out-of-plane hoppings are negligibly small\cite{mackenzie1996,shen2000} and we shall consider the 2D limit hereafter. 

In the left panel of Fig. \ref{fig:edgestates} we show the Fermi-surfaces. There are three Fermi-surfaces: two around $(k_x,k_y)=(0,0)$ and one around $(k_x,k_y)=(\pi,\pi).$ The two quasi-1D Fermi-surfaces from $d_{xz}$ and $d_{yz}$ orbitals do not touch as long as $\lambda\neq 0.$ The inner quasi-1D Fermi-surface is a hole pocket, and the outer Fermi-surface is an electron-pocket. The round quasi-2D Fermi-surface arises from the $d_{xy}$ orbital which we assume is completely decoupled from the quasi-1D orbitals at the single-particle level and in the 2D limit\footnote{Weak hybridization between quasi-2D and quasi-1D orbitals would occur when out-of-plane hoppings are considered.}. 
We have also left out the spin-orbit coupling from this Bloch Hamiltonian description. It is expected that the spin-orbit coupling scale in Sr$_2$RuO$_4$ is appreciable, and that it affects the orbital character of the states on the Fermi surfaces, primarily near the intersections between the quasi-1D and quasi-2D Fermi surfaces\cite{liu,sawatzky2008}. The main effect of the spin-orbit coupling will be to determine the dominant superconducting pairing instabilities on each Fermi surface. As such, since this article is agnostic toward which order parameter dominates we will not consider the corrections due to spin orbit coupling much further. We will thus make the same assumption as RKK, i.e.  the inter-orbital hybridization is more important than the spin-orbit coupling, and as such we have spin-rotation symmetry. We do note that if the spin-orbit coupling is too strong it might destabilize the MBS zero-modes on the linked dislocation lines considered below, if spin-rotation symmetry is strongly broken. However, it is likely that before this happened the dominant order parameter would be modified significantly anyway to a (possibly) different topological class. As such, we will leave the consideration of the non-trivial effects of spin-orbit coupling in Sr$_2$RuO$_4$ to future work.

We now want to consider the properties of the superconducting state of Sr$_2$RuO$_4.$ When we consider the quasi-2D band we will assume triplet $p_x+i p_y$ pairing (possibly induced via proximity coupling to the quasi-1D bands\cite{rkk2010}). For values of the Fermi-level which lie within the quasi-2D band (which is expected in experiments) this means that the system will be a weak 3D TSC with primary index ${\bf n}=(0,0,1).$  For the quasi-1D bands we assume either a topologically trivial pairing\footnote{We will not specify what the pairing might be, only that $\bv$ vanishes.}  or the RKK pairing, which we now describe. Since the quasi-1D and quasi-2D orbitals are assumed to be approximately decoupled, at least at the single-particle level, we can separate off the quasi-2D band and write a reduced two-orbital model for the quasi-1D orbitals:
\begin{eqnarray}
H({\bf k})=\left(\begin{array}{cc}\epsilon_{xz}({\bf k}) & \Lambda({\bf k}) \\ \Lambda({\bf k})& \epsilon_{yz}({\bf k}) \end{array}\right)\otimes {\mathbb I}_{spin}
\end{eqnarray}\noindent with $\lambda=0.1t.$
The superconducting pairing that RKK propose is spin-triplet and intra-orbital. The pairing functions of orbital $\alpha$ for this chiral superconducting state are
\begin{eqnarray}
\Delta_\alpha &=& i {\bf d}_\alpha({\bf k})\cdot {\vec\sigma} \sigma^y, \;\;\; \alpha=1,2\\
{\bf d}_1&=&\hat{z}\Delta_0 \sin k_x \cos k_y,\\
{\bf d}_2&=&i\hat{z}\Delta_0 \sin k_y \cos k_x,
\end{eqnarray}
where the direction of ${\bf d}_\alpha$ and the relative phase between ${\bf d}_1$ and ${\bf d}_2$  are determined by the inter-orbital hybridization and spin-orbit coupling. This pairing term establishes a non-trivial secondary weak invariant $\bv=(1,1,0).$

When considering all of the orbitals there are a few possible scenarios for the pairing in Sr$_2$RuO$_4,$ but let us limit ourselves to two main cases: (i) the pairing is dominated by the quasi-2D orbital and is the chiral, topological $p_x+ip_y$ state so that ${\bf n}\neq 0$ but $\bv=0$ or (ii) the pairing is dominated by the quasi-1D orbitals which are in the RKK state and a (probably weak) $p_x+ip_y$ state is induced on the quasi-2D orbital, i.e. ${\bf n}\neq 0$ and $\bv\neq 0.$   It is these two cases which we aim to topologically distinguish in this article.

\emph{Properties of boundary states}:
The RKK superconducting pairing term winds around the two quasi-1D Fermi surfaces with the same chirality, but since they have opposite charge character, they contribute oppositely to the winding number yielding a vanishing Chern number. 
However, in a clean system with (even approximate) translation invariance there will be edge states located near, say $k_x=0$ and $k_x=\pi$ if, for example we put the system on a cylinder with open boundary conditions in the y-direction and periodic boundary conditions in the x-direction. The energy spectrum for such an open boundary system is shown in the right panel of Fig. \ref{fig:edgestates} with clear low-energy modes near $k_x=0,\pi$ which develop zero modes exactly at these k-points. This figure assumes the pairing scenario (i) and  the boundary states exist because of a non-trivial primary weak index ${\bf{n}}=(0,0,1)$, and a secondary weak index $\bv=(1,1,0)$ due to the RKK pairing.  Even though ${\bf n},$ and other quantities, should be doubled when the spin degeneracy is taken into account, it has no qualitative effect on most of the properties of lattice dislocations discussed below when the effects of spin-orbit coupling are weak and we have approximate spin-rotation invariance. Thus the quasi-1D bands contribute gapless boundary states, albeit non-chiral modes. The energy gaps in the figure have been exaggerated from what one would expect in a real experiment to illustrate the important features.

 If the $p_x+ip_y$ state on the quasi-2D Fermi surface is generated via proximity coupling (or exists independently of the quasi-1D bands as in scenario (i)) then there will be an overall chirality of the boundary states, e.g. two chiral modes at $k_x=0$ and one anti-chiral mode at $k_x=\pi.$  If the location of the surface states in momentum space can be resolved via ARPES then, if the RKK pairing is not present, we would not expect any gapless surface states at $k_x=\pi$ ($k_y=\pi$) for edges with normal vectors in the $\hat{y}$ ($\hat{x}$) direction. This is one distinguishing feature of these two pairing scenarios. If the gap induced on the quasi-2D band is very weak then the chiral boundary states associated to this gap will not be very well localized and might hybridize with the low-energy modes on other boundaries. In this case, there will be an energy gap for the chiral modes and the gapless boundary states will be distinctly non-chiral.  This is perhaps a more clear signature, and is one feature that RKK emphasized, however from our analysis above, the boundary state distinction persists between the two pairing scenarios even if the $p_x+ip_y$ gap is not extremely small. 

\emph{Properties of dislocations and linked dislocation lines}:
In addition to the surface state properties, weak topological indices have important consequences for the properties of crystal  dislocations\cite{ran2009,ran2010,teo2010}. For the 3D crystal we are considering, a lattice dislocation is a line defect around which the ions are displaced by an integer valued vector ${\bf b}$ (in the lattice basis)  known as the Burgers vector. A dislocation is described by ${\bf b}$ and the integer-valued tangent vector  ${\bf l}$. The relative orientation of ${\bf b}$ and ${\bf l}$ determines the type of dislocation: edge (${\bf l}\cdot{\bf b}=0$), screw (${\bf l}$ parallel to ${\bf b})$, and mixed (${\bf l}$ neither parallel nor perpendicular to ${\bf b}$). While ${\bf b}$ is a topological property of a dislocation line, $\bf{l}$ 
(namely the dislocation-type) is not.

\begin{figure}[t]
    \begin{center}
        \includegraphics[width=3in]{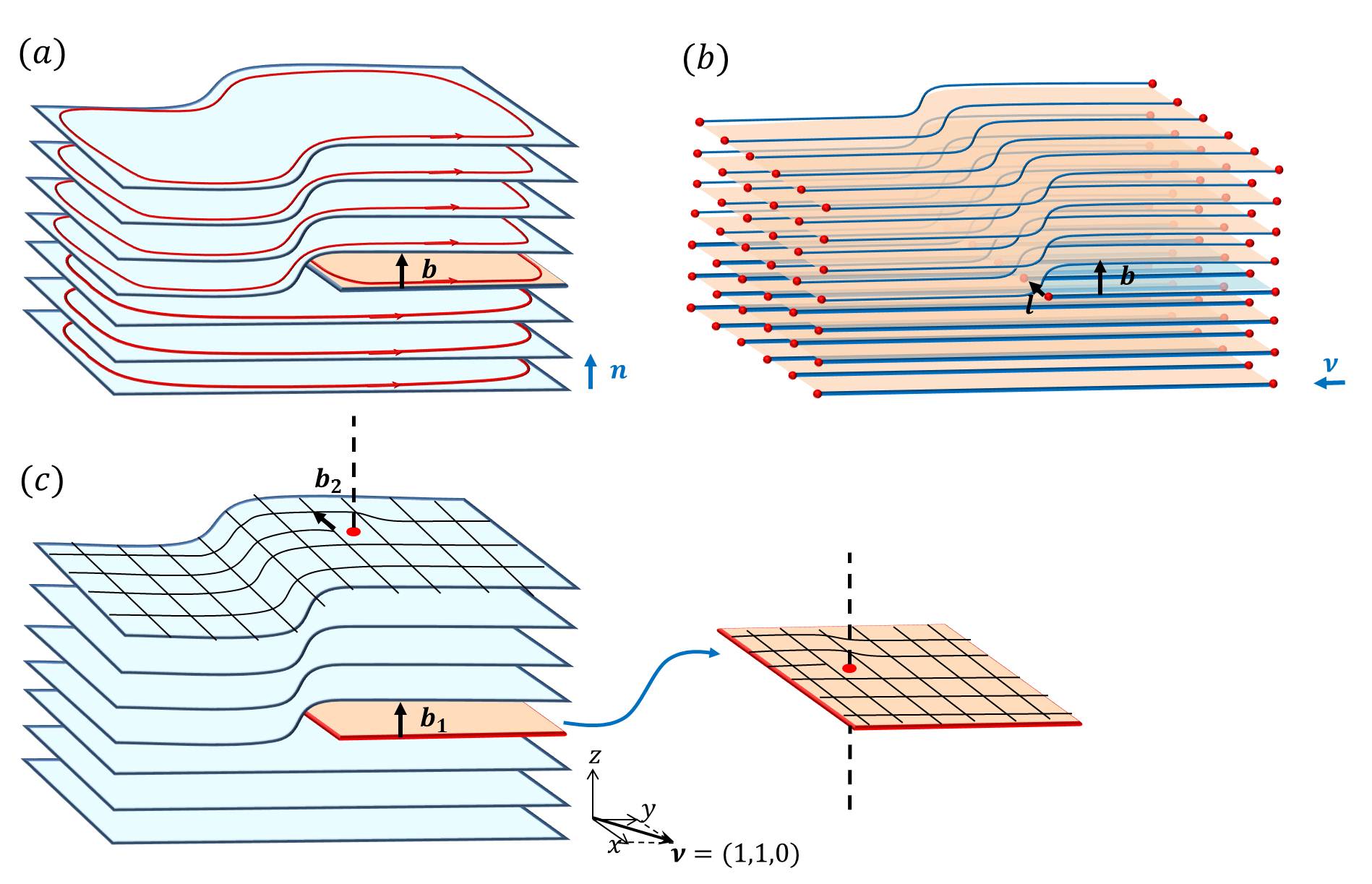}
    \end{center}
    \caption{(a) Illustration of an edge dislocation line with Burgers vector ${\bf b}$ in a TSC consisting of decoupled layers stacked along the direction of topological index vector ${\bf n}$. 
    (b) Illustration of an edge dislocation line with Burgest vector ${\bf b}$ and direction ${\bf l}$ in a TSC consisting of decoupled 1D wires along the direction of secondary weak topological index vector $\bv$. 
    The red dots at the end of wires represent Majorana zero modes.
    (c) Illustration of the Majorana zero modes induced by linking of two edge dislocations with ${\bf b}_1=\hat{z}$ and ${\bf b}_2=-\hat{x}$ in the decoupled plane limit.
    There is a Majorana zero mode at each dislocation line, indicated by the red dot and red line.}
    \label{fig2}
\end{figure}

Both the primary and secondary weak topological indices can be probed by dislocations. The primary weak indices ${\bf n}$ in class D lead to \begin{equation}N_1={\bf n\cdot b}\end{equation} branches of chiral Majorana modes propagating along the dislocation\cite{ran2010}. Note that $N_1$ is independent of the dislocation tangent vector ${\bf l},$ and is thus topologically protected. The sign of the integer $N_1$ determines the chirality of the localized state, which is defined with respect to ${\bf l}$.  This fact can be easily understood for an edge dislocation with ${\bf n}$ perpendicular to ${\bf l}$, in which case the dislocation can be obtained by adding an additional layer of chiral 2D TSC on one side of the dislocation line, as is illustrated in Fig. \ref{fig2} (a), i.e. we just imagine jamming an extra partial-plane of a 2D strong chiral TSC into the layered system. 

The secondary weak indices $\bv$ lead to non-chiral 1D propagating Majorana modes on the dislocation line if \begin{equation}N_2\equiv\left({\bf b}\times{\bf l}\right)\cdot\bv=1 \mod 2.\end{equation} The modes determined by $N_2$ are like the ``weak" analog of the ``strong" modes determined by $N_1.$ This fact can be easily understood using a similar picture for an edge dislocation,  in which case the dislocation can be obtained by adding an additional layer of a \emph{weak}  2D TSC on one side of the dislocation line, as is illustrated in Fig. \ref{fig2} (b), i.e. we imagine jamming an extra partial-plane of a 2D weak TSC into the layered system. 

For the secondary weak invariant, the dislocation line is thus like the edge of a weak TSC and requires translation symmetry to have protected modes.
We can see this because of the dependence of $N_2$ on the variable direction ${\bf l},$ which indicates that topological stability will require an additional symmetry which in this case is translation symmetry along the dislocation (\emph{i.e.} the direction ${\bf l}$ cannot change along the dislocation line). The non-chiral Majorana propagating modes are protected by translation symmetry along the dislocation, since its left and right moving branches are around $k=0$ and $k=\pi$ ($k$ is the momentum parallel to the translation invariant dislocation line), which cannot be coupled without breaking translation symmetry. Also we see that $N_2$ can be nontrivial only for edge dislocations, i.e. ${\bf b}\times {\bf l}\neq 0$ must be satisfied.

 In the topologically equivalent decoupled chain limit (which is appropriate for a system with $\bv\neq 0$), the dislocation bound states can be understood intuitively, as is illustrated in Fig. \ref{fig2} (b). Decoupled 1D Majorana chains along the $\bv$ direction terminate at the dislocation line and the MBS at their end points couple to form the 1D non-chiral Majorana edge state. It is thus intuitive to take $N_1$ to be akin to a  ``strong" \emph{dislocation} invariant and $N_2$ to be a ``weak" \emph{dislocation} invariant. The weak invariant $N_2$ requires the additional translation symmetry along the dislocation line to be protected. 

\begin{figure}[t]
    \begin{center}
        \includegraphics[width=2.3in]{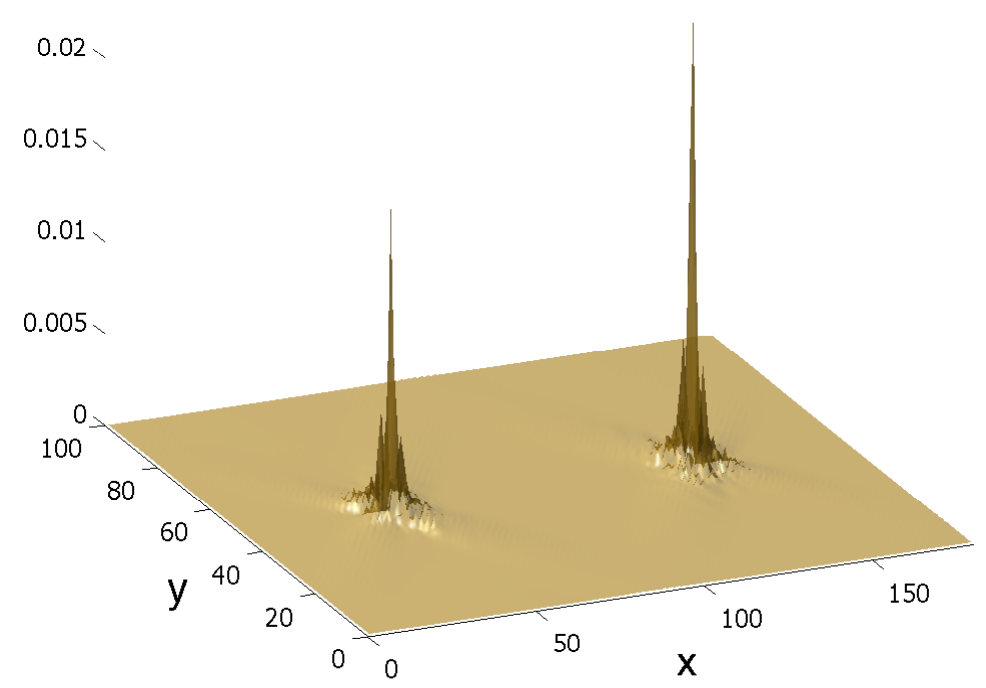}
    \end{center}
    \caption{Spatial profile of the sum of the probability densities  for the two lowest energy (closest to zero energy)  bound states at the ends of a dislocation that stretches from $x=40$ to $x=120$ on the line $y=50$. The total lattice size is $160\times 100$. The asymmetry in the density profile is due to finite-size error introduced by discretization.}
    \label{fig:dislocation}
\end{figure}

Despite the fact that our arguments are based on the decoupled layer/wire limit, the topological protection remains as long as the bulk gap is not closed. Thus, we can move away from the decoupled limit and the low-energy bound states will remain stable. We illustrate this numerically in Fig. \ref{fig:dislocation} where we plot the sum of the  probability densities for the two zero energy modes which are localized on the two dislocation lines that bound an extra partial-plane of atoms. To perform the calculation we only considered the quasi-1D RKK sector, since the quasi-2D band is decoupled and we wanted to test the secondary weak invariant. We used the same parameters as in Fig. 2 where the layers/wires are certainly coupled. We inserted an edge dislocation where the Burgers' vector of the dislocation is ${\bf{b}}=(0,1,0)$ and the tangent vector is ${\bf l}=(0,0,1)$ which yields $N_2=(\hat{y}\times \hat{z})\cdot(\hat{x}+\hat{y})=1.$ We used exact diagonalization on a system with periodic boundary conditions in all three directions to extract the wavefunctions of the two zero modes. We plot the sum of their probability densities (restricted to the $xy$-plane) in Fig. \ref{fig:dislocation}. One can clearly see the exponentially localized bound states on each dislocation line.

One additional effect that has thus far gone unnoticed is the property of linked dislocation rings. 
Along a finite-length dislocation ring in a system with nontrivial topological invariants $N_1$ and/or $N_2$, the Majorana fermion energy spectrum is discrete, and the boundary conditions of the fermions around the ring determine whether there is an exact zero-energy Majorana mode (c.f. Ref. \onlinecite{stoneroy2004}). Interestingly, the boundary condition around a dislocation ring depends on its linking with 
other dislocation rings and 
with flux/vortex lines. To illustrate, consider the RKK model with $\bv=(1,1,0)$ and consider two edge dislocation lines: one which is a circle in the $xy$-plane with ${\bf b}_1=(0,0,1),$ and one with ${\bf b}_2=(1,0,0)$ aligned along the $\hat{z}$ direction so that ${\bf l}=\hat{z}.$  If these two dislocations are not linked, the MBS along the $xy$-plane dislocation loop has a finite size gap with no exact zero mode because the boundary conditions are (effectively) anti-periodic due to a Berry phase effect\cite{stoneroy2004}. This can be shown in the decoupled-layer limit as in Fig. \ref{fig2}(c), in which case the in-plane dislocation circle is the boundary of a single-layer disk, and a finite gap of order $1/R$ (with $R$ the radius of the circle) is present since the boundary conditions are (effectively) anti-periodic. 

In contrast, when the circle encloses the other dislocation line along $\hat{z}$ direction, the effect is to introduce an edge dislocation with Burgers vector ${\bf b}_2$ in the disk, which introduces an extra translation phase for the fermion modes on the $xy$-plane dislocation loop enclosing the threaded dislocation line. When the condition  \begin{equation}N_0\equiv N_L\left({\bf b}_1\times {\bf b}_2\right)\cdot \bv = 1 \mod 2,\end{equation} where $N_L$ is the linking number of the two edge dislocations, is met, the boundary conditions are shifted exactly back to effectively periodic, and a MBS will appear. Since MBS have to come in pairs, there must also be a MBS along the other dislocation which is threaded through the disk\footnote{Couplings between two Majorana zero-modes can only generate a exponentially small gap of order of $e^{-R/\xi}$, where $\xi$ is the correlation length of the superconductor.}. In fact, if the dislocation line along $\hat{z}$ is glued to form a closed loop then it will also clearly receive a translation phase which will convert the boundary conditions to effectively periodic exactly when the same condition is met. Since any generic superconductors can be adiabatically deformed to the decoupled layer limit (due to the absence of strong invariant in 3D), the number of MBS on linking dislocations can be determined generically from this argument. Importantly, the dependence on ${\bf l}$ has dropped out which means that $N_0$ is topological and does not require the addition of translation symmetry along the dislocations. Thus, we see that a crucial consequence of the secondary weak invariant is the determination of bound states on linked dislocations. This is one of the main results of this article, and it provides a mechanism to generate stable MBS on defects even in superconductors with an even-integer or vanishing Chern number. If we had left $\bv$ as an anti-symmetric tensor this invariant would simply be the contraction of the tensor with the Burgers' vectors of both dislocation lines.

While the primary weak invariant $N_1$ has no effect in the linking of two dislocations, it does determine the MBS when linking occurs between dislocation lines and flux/vortex lines. When a superconducting vortex ring is linked with a dislocation ring, the boundary condition for the Majorana fermion along the dislocation line will change. For odd $N_1$ such a boundary condition change results in a single MBS on the dislocation line, and another one on the vortex line. The existence of a MBS on the dislocation line is determined by \begin{equation}\tilde N_0\equiv\tilde N_L N_1 \mod 2,\end{equation} where $\tilde N_L$ is the linking number between a dislocation line and a vortex line.

\emph{Discussion and Physical Consequences of Dislocation Bound States}: 
So far we have mentioned how one might distinguish the two pairing scenarios using the low-energy boundary modes. Now let us indicate several more distinguishing features determined from our dislocation analysis above.

(1) Due to the nontrivial primary weak topological invariant ${\bf n}$, the dislocations with Burgers vector ${\bf b}=\hat{z}$ have chiral Majorana fermion modes. Thus, in scenario (i)  each such dislocation carries a (localized) ``persistent" chiral energy current $I_E=\frac{\pi k_B^2T^2}{24\hbar}$ at finite temperature\cite{kane1997}.
However, we should clarify that this "persistent energy current" carried by a chiral Majorana mode is not in contradiction with the fundamental principles of thermal transport (such as heat can only flow in non-equilibrium) because the number of left and right moving Majorana modes is always the same in any physical system. For our case this implies that since in a physical crystal dislocation lines must come in pairs if they extend between two boundaries, or form loops if the do not, there is no way to extract an energy transport current without perturbing the system away from equilibrium.  Locally there is an energy current on every dislocation even without a thermal gradient, however a real transport experiment is sensitive to the global energy current and the currents of the dislocations and anti-dislocations and dislocation loops will globally cancel. 
With a random distribution of dislocations in the system, a net chiral energy flow will not be observed when a temperature gradient is applied. However, the chiral energy current along random dislocations will contribute a thermal conductivity that is proportional to the dislocation density. Furthermore, it is possible to have a strained system with imbalanced dislocation lines which could conduct a net energy current since the strain will cause an inhomogeneous dislocation density. We should note that similar energy flows might also be observed on vortex lines in TSCs that have bound low-energy chiral modes\cite{silaev1,silaev2}. 

Compared to the thermal current carried by the edge states which is easily overwhelmed by bulk thermal conduction, the dislocation current can be a bulk effect that remains finite in the thermodynamic limit, and this would be a signature of scenario (i) where the $p_x+ip_y$ quasi-2D pairing dominates. In scenario (ii), if the $p_x+ip_y$ gap is weak then the chiral modes on the dislocations will not be localized and will hybridize with other low-energy modes and annihilate. Thus, we generically would not expect chiral energy currents on dislocations unless the the $p_x+ip_y$ gap is strong. 


(2) Another distinguishing feature arises for linked defects. For an edge dislocation ring in the $xy$-plane (with Burgers vector ${\bf b}_1=\hat{z}$) and a second edge dislocation along the $z$-axis 
threading the ring (with Burgers vector ${\bf b}_2$ in-plane), a Majorana zero mode on the second dislocation may be observable by scanning tunneling microscopy (STM). In particular, when the first dislocation is at a crystal surface, it will be a disk-shaped plateau on the surface, threaded by the second dislocation line and be easy to locate using, for example, TEM. If linked dislocations exhibit the zero-energy MBS it is an immediate smoking gun indication for the RKK pairing, i.e. scenario (ii). 

Given the same set up for the surface dislocation we can also apply a magnetic field. For scenario (i) the STM signal will exhibit a clear even-odd dependence based on the number of vortices piercing the dislocation ring plateau. When the number of vortices is odd then there should be a zero-energy bound state on both the vortex and the dislocation loop, and when the number of vortices is even then there will not be a zero mode. Thus, the even-odd effect would be a clear signature of scenario (i). However, if the proximity-induced $p_x+ip_y$ gap in scenario (ii) becomes large then this effect would also be present in that case. However, since we expect the gap to be weak then the zero-mode on the vortex and that on the dislocation loop will hybridize and open a gap, thus presenting the observation of any zero-modes in STM.

These two possibilities yield useful experimental proposals for how to distinguish the two pairing scenarios. To some extent, since the proposals are based on weak topological invariants, then they should be protected by translation symmetry. Even though clean crystal samples of Sr$_2$RuO$_4$ are available there will always be some disorder which breaks the translation symmetry. Thus, some comments on the stability of our predictions in the presence of disorder are necessary, although this topic is an active area of research and we will leave a full discussion for future work.

For the primary weak invariant, i.e. either pairing scenario (i), or scenario (ii) with a strong induced $p_x+ip_y$ quasi-2D gap, the low-energy bound states on a dislocation line will be chiral, and robust to the addition of disorder as long as no low-energy, delocalized channels couple to the modes. We thus expect the modes to be stable for weak to moderate strength disorder. In order for the modes to be destroyed there must be some mechanism for them to leak away from their dislocation line and couple to modes on a separate dislocation line in the bulk or on the surface. There could be some local fluctuation of the disorder potential that might destabilize the modes in a local region, or the disorder could be strong enough to close the bulk/mobility gap which would signal a disorder-driven bulk topological phase transition. The stability is essentially the same if we consider the boundary states themselves which are also chiral, and for zero-energy bound states on a dislocation line linked with a vortex as long as $\tilde N_0$ is non-trivial.

 For the secondary weak invariant there will exist non-chiral modes on dislocation lines and non-chiral boundary modes. These modes are not as stable as the modes due to the primary invariant as they can be gapped/localized locally at the dislocation line (or boundary)  itself. This can happen from local disorder at or near the dislocation line which can localize the low-energy modes. Even non-random, but translation-symmetry breaking perturbations could localize the modes. We thus expect these modes to be stable only in the case of weak disorder. 

As we have emphasized the secondary weak invariant also has another consequence. That is, for the secondary weak invariant there will exist zero-energy bound states on linked dislocation lines. These modes are insensitive to disorder on the dislocation itself, but can be destroyed if they come too close, or couple through delocalized modes, to other zero-energy states. These modes are thus the most stable consequence of bound states due to the secondary weak invariant and we expect them to be stable for weak to moderate disorder.

In conclusion, we have observed that the RKK pairing induces a non-trivial secondary weak invariant which distinguishes it from the primary weak invariant of the chiral $p_x+ip_y$ state. We then showed that this secondary weak invariant has several interesting consequences which can help to experimentally distinguish the two pairing scenarios, the most notable of which is the presence of Majorana bound states on linked dislocation lines. For future work it will be exciting to consider the effects of strong spin-orbit coupling and disorder. 


%
%

After the completion of this work we noticed a recent preprint with similar themes albeit a different focus\cite{ueno2013}.
{\bf Acknowledgement}: We thank S. A. Kivelson, Y. Liu, A. Mackenzie, S. Raghu, and M. Silaev for helpful discussions. This work is supported in part by  the US Department of Energy under contract DE-FG02-07ER46453 (TLH) and ONR award N0014-12-1-0935 (TLH), the Tsinghua Startup Funds and the National Thousand Young Talents Program of China (HY), and the National
Science Foundation through the grant No. DMR-1151786
(XLQ).

\end{document}